\documentclass[longbibliography,epjST]{svjour}
\usepackage{graphics}
\usepackage{physics}
\begin{document}
\title{Spatial coupling of quantum-anomalous-Hall and chiral-Majorana modes}
\author{Javier Osca\inst{1}\and Marc Alomar\inst{1} \and Lloren\c{c} Serra \inst{1,2}\fnmsep\thanks{\email{llorens.serra@uib.es}}  }
\institute{Institute for Cross-Disciplinary Physics and Complex Systems
IFISC (CSIC-UIB), E-07122 Palma, Spain 
\and Physics Department, University of the Balearic Islands, E-07122 Palma, Spain
}
\abstract{
We calculate density and current spatial distributions of a 2D model junction between a normal QAH contact and a superconducting QAH region hosting propagating (chiral) Majorana modes. We use a simplified Hamiltonian describing the spatial coupling of the modes
on each side of the junction,  as well as the related junction conductance. We study how this coupling is affected 
by orbital effects caused by an external magnetic field. 
} 
\maketitle
\section{Introduction}
\label{intro}
Majorana states in Condensed Matter have been a hot topic for a few years
now \cite{Nayak,Qi,Alicea,StanescuREV,Beenakker,Franz,Elliott,Aguado,Lutrev,Lutchyn,Oreg}. 
Different experiments have been carried out in order to demonstrate the actual existence of such topological states. 
Majorana modes are characterized by being chargeless and spinless edge states, hence
most of the experiments aiming at their detection are based on
identifying characteristic signatures on the electrical conductance of devices
attached to them \cite{Mourik,HaoZ,Deng,Das,He294}.
To obtain Majorana states one needs 
the presence of superconductivity, therefore the typical scenario usually requires a contact between a normal lead and a hybrid proximity-coupled semiconductor-superconductor. As topological states, Majorana modes are separated by an energy gap that protects them from other normal states and local sources of noise, a robustness that might allow the use of such states for topological quantum computing.

In many ways Majoranas can be understood as non-local split Fermions. In this sense there are two kinds of Majorana states: non-propagating Majorana states appearing at the ends of  (quasi) 1D nanowires and propagating chiral Majorana states formed along the edges of 2D-like hybrid structures. In this work we will focus on the second kind. We refer, more specifically, to devices similar to those of Refs.\
\cite{He294,Qi06,Qi10,Chung,Wang15,Lian16,Kalad}
consisting of a quantum Hall (QH) or quantum anomalous Hall (QAH) insulator proximity coupled with a superconductor (QAH+S). 
In particular, we will consider a simple model of QAH+S that does not need the presence of external magnetic fields. In this kind of systems, chiral Majorana modes propagate along the edges in a clockwise or anticlockwise manner (depending on device parameters) for finite systems. 

An open infinite nanowire like the one depicted in the inset of Fig.\ \ref{F1} may hold two pairs of counterpropagating Majorana channels, one pair at each edge of the device. In general, it has been reported that each chiral Majorana channel contributes $0.5 e^2 / h$ to the linear conductance of a device. However, in this work we will show that for the infinite nanowire with only one normal contact the conductance remains $e^2/h$ independently of the number of active Majorana modes (one or two), even with a finite transmission probability to the Majorana channel of $\approx0.5$. The reason for this is that we consider a single normal contact connected to a semi-infinite Majorana device, instead of the two usual contacts in a normal-superconductor-normal arrangement. When only one normal contact (left) is present, only half of the possible Majorana channels are active, the  outgoing ones. Ingoing Majorana modes into the junction would necessarily require a second (right) normal contact and therefore they are not contributing in our arrangement.

We use a method based on the evaluation of the (complex) wave numbers allowed on each side of the junction and giving the detailed spatial distribution patterns of density and currents.
In addition, we study how the spatial distribution of the Majorana modes is affected by magnetic orbital effects, on top of the already present QAH physics. We show how the spatial coupling between Majorana and non-Majorana states at both sides of the junction modifies the transmission and reflection processes, and thus also the conductance.
This article is divided in five parts. Sections \ref{sec:1} and \ref{sec:2} present the model and the method of resolution to determine ingoing and outgoing modes of the junction. Next, in  Secs.\ \ref{sec:3} and \ref{sec:4} we present the results without and with orbital effects of the magnetic field, respectively. Finally, a summary and outlook of the work is given in Sec.\ \ref{sec:5}.

\section{Model}
\label{sec:1}

Our main objective is to study the distribution of currents and the conductance present in a N-(QAH+S) junction where chiral Majoranas may be present.
We start using a simplified model of QAH+S Hamiltonian similar to the one 
devised in Refs.\ \cite{Qi06,Qi10},
\begin{equation}
h_{\it BdG}({\bf p})=m({\bf p}) \sigma_z - \alpha\,(p_x \sigma_y - p_y \sigma_x)\tau_z + \Delta(x)\,\tau_+ + \Delta(x)^*\, \tau_-\; ,
\label{E1}
\end{equation}
where $m({\bf p})=m_0+m_1 {\bf p}^2$, with $m_0$ and $m_1$ known material parameters. 
As usual, the $\sigma$'s and $\tau$'s represent Pauli matrices for spin and isospin, respectively.
We will consider $\alpha$ a known parameter related with the quasi-particle mass governing the shape of the Dirac cone for energies near its apex. In this work we set $\alpha\equiv 1$ as our unit for practical reasons. We will assume superconductivity achieved by proximity coupling between the QAH semiconductor and a metallic superconductor. The union between a superconducting and non superconducting region
will be achieved through the spatial variation of the superconductor coupling constant $\Delta(x)$. 
The numerical results of this work will be presented in natural units of the problem, i.e., 
taking $2m_1$, $\hbar$ and $\alpha$ as unit values.
That is, our length and energy units are $L_U\equiv L_{so}=2m_1\hbar^2/\alpha$ and 
$E_U=\alpha^2 /2m_1\hbar^2$.

\begin{figure}
\center
\resizebox{0.75\columnwidth}{!}{%
  \includegraphics{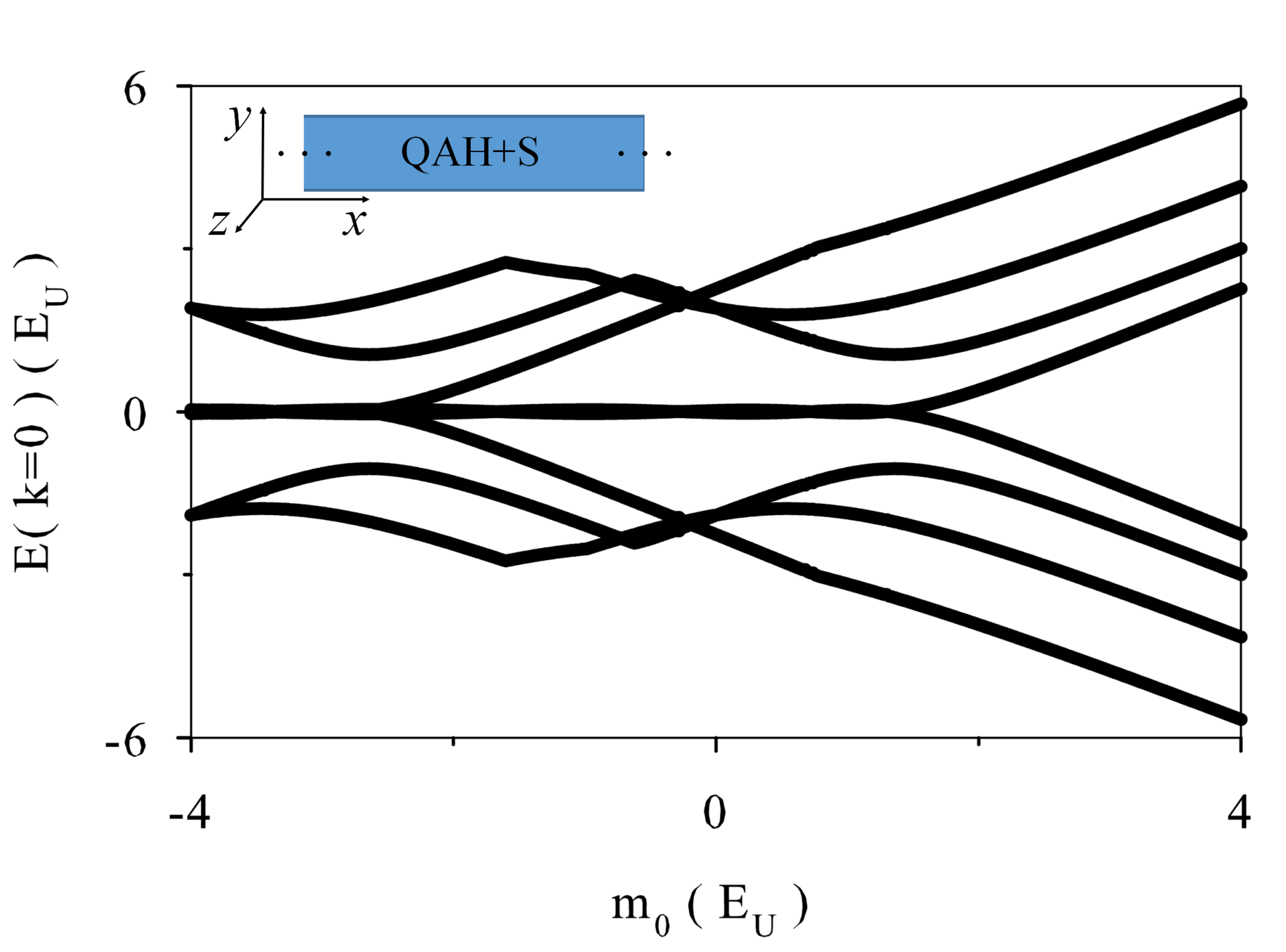}}
\caption{$E(k=0)$ as a function of the material parameter $m_0$ for a QAH slab of $L_y=5 L_U$ proximity coupled with a superconductor
yielding strength $\Delta=2 E_U$. 
A sketch of the infinite system
used for this band structure calculation is given in the inset.
Notice the phase transitions at $m_0\approx\pm\Delta$, as indicated by the presence of zero modes. }
\label{F1}       
\end{figure}

This model provides two phase boundaries with a critical value of the $m_0$ parameter, 
$m_{0}^{(c)}=\pm |\Delta|$. For large positive values of $m_0$ the device will be in a trivial phase while for large negative ones a phase of Chern number ${\cal C}=2$ will arise with two chiral Majoranas attached to each edge of the device. For intermediate values of $m_0$, between the two phase boundaries, there is a single Majorana phase of Chern number one (see Fig.\ \ref{F1}).
The phase-transition boundaries may differ slighty from these values due to 
the transversal confinement, in a similar manner as in non-chiral Majorana nanowires \cite{Osca2015b}. 
Of course, the effect of the transversal confinement becomes negligible in wide enough wires.

The presence of the Majorana modes is signaled by a pair of topological bands at wavenumber $k=0$ for the translationally invariant (infinite) wire. In Fig. \ref{F1} this can be seen with a plot of the energy $E(k=0)$ as a function of $m_0$. The presence of zero-energy modes indicate the Majorana phases, in good
agreement with the expected critical values. The bulk-edge correspondence principle 
ensures that the critical value $m_0^{(c)}$ also indicates when chiral Majoranas 
will appear in a semi-infinite nanowire or in the superconducting region of the N-(QAH+S) junction studied in this work.

\section{Method}
\label{sec:2}

\begin{figure}
\center
\resizebox{0.75\columnwidth}{!}{%
  \includegraphics{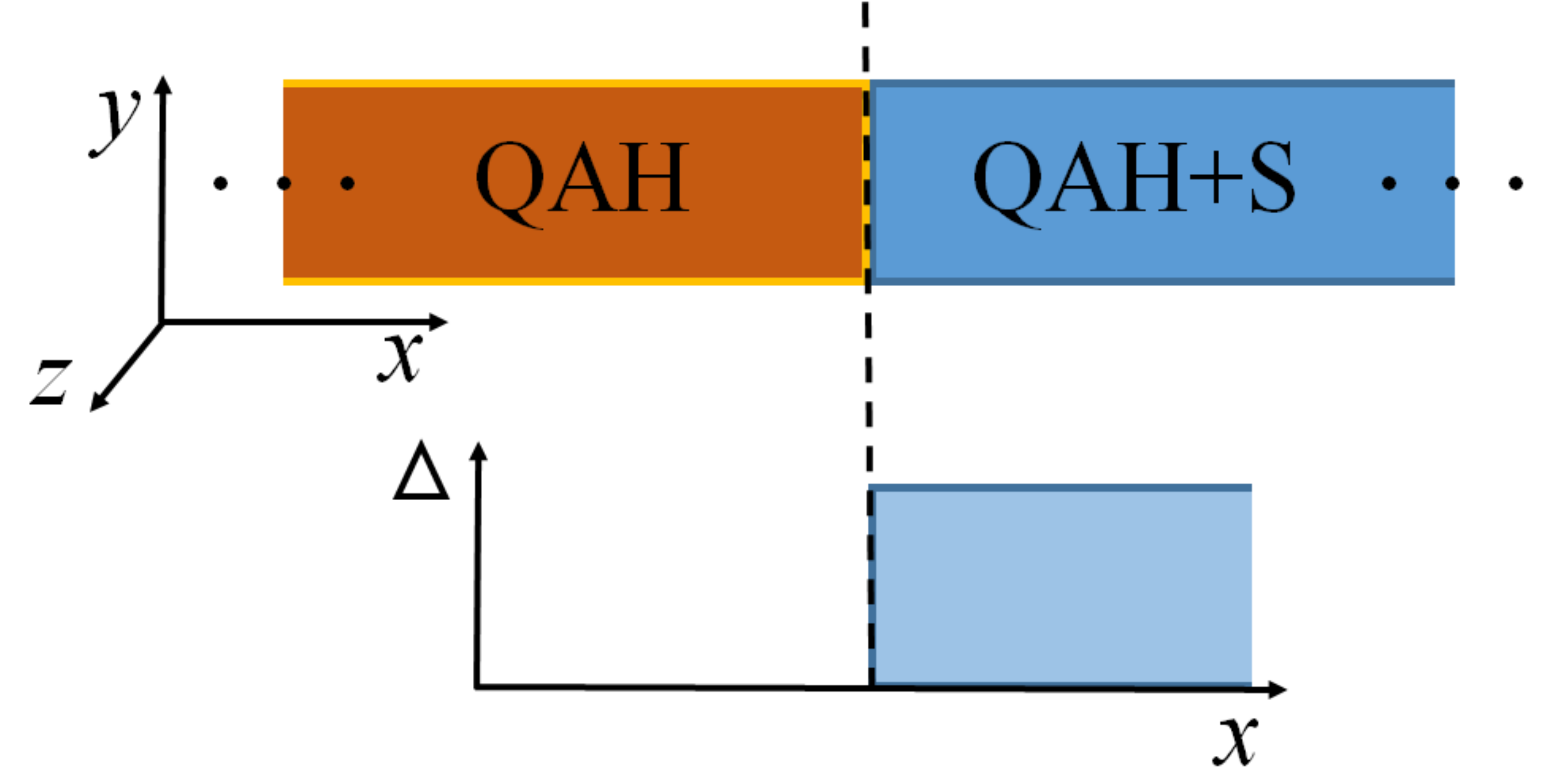}}
\caption{Graphical description of the nanowire junction considered in this work, an infinite QAH slab with half of the slab proximity coupled with a superconductor. The junction interface separates normal and superconducting regions. On the left side there is a normal QAH region while on the right side
there is a hybrid QAH+superconducting region with a non-zero $\Delta$.}
\label{F2}       
\end{figure}

We want to calculate the distribution of currents for a junction between a normal QAH material and a material of the same kind proximity coupled with a superconductor (see Fig.\ \ref{F2} for a graphical representation of the device). The numerical method was already used by us to calculate local currents and
conductance in N-S junctions for non-chiral Majoranas in Refs.\ \cite{Osca2017,Osca2017b},
with some technical differences as briefly explained below.

The overall idea is that of a matching method for two different sets of asymptotic solutions, for  a given energy $E$, one for each side of the junction and characterized by a $k$ wave number,
$\Psi_{k}(x,y,\eta_\sigma,\eta_\tau) =\phi_{k}(y,\eta_\sigma,\eta_\tau) e^{i k x}$. These asymptotic solutions for the left and right contacts are assumed to be known for a large-enough set of wave numbers, with 
$k$ being either real (propagating) or complex (evanescent) \cite{Serra13}.
The full solution for  the left and right sides of the junction ($c=L,R$) is given by a superposition
of the corresponding set of modes,
\begin{equation}
\label{eq2}
\Psi^{(c)} (x,y,\eta_\sigma,\eta_\tau)=\sum_{k} d_{k}^{(c)}	\, {e}^{i k x}\, \phi_{k}(y,\eta_\sigma,\eta_\tau)\,.
\end{equation}
The wavenumbers and the transverse  eigenstates can be obtained numerically as solutions of the BdG Hamiltonian for each contact, where  $ \sum_{\eta_\sigma \eta_\tau} \int{ dy\, \phi_{k}}=1$. The coefficients $d_{k}^{(c)}$ that determine the strength of each channel are obtained from the matching algorithm \cite{Osca2017,Osca2017b}.

The distribution of currents is calculated from the wave functions given by Eq.\ (\ref{eq2}).
We consider three different kinds of densites $\rho_a(x,y)$ and current $\vec{j}_a(x,y)$, where subindex  $a$ may be $a=qp, c, s$ for quasiparticle, charge, and spin, respectively.  Quasi-particle distributions are given by 
\begin{eqnarray}
\label{E2a}
\rho_{qp}(x,y) &=& \Psi^*(x,y)\Psi(x,y) \; ,\\
\vec{j}_{qp}(x,y) &=& \Re\left[\, \Psi^*(x,y)\, \hat{\vec{v}}_{qp}\, \Psi(x,y)\, \right]\;,
\label{E2}
\end{eqnarray}
where the velocities are given by  $\hat{v}_{qp,x}=\partial \mathcal{H}/\partial p_x$ and  $\hat{v}_{qp,y}=\partial \mathcal{H}/\partial p_y$. Quasiparticle density and current fulfill a  continuity equation $\partial \rho_{qp}(x,y) / \partial t=\nabla\cdot\vec{j}_{qp}(x,y) $ because the model has no sources or sinks of quasiparticles. 
 With the Hamiltonian of Eq.\ (\ref{E1}) it is,
\begin{eqnarray}
\hat{v}_{qp,x} &=& -i \hbar 2 m_1 \partial_x \sigma_z - \frac{\alpha}{\hbar} \sigma_y \tau_z\;, \label{E3} \\
\hat{v}_{qp,y} &=& -i \hbar 2 m_1 \partial_y \sigma_z + \frac{\alpha}{\hbar} \sigma_x \tau_z\;. \label{E4}
\end{eqnarray}

Substitution of Eqs.\ (\ref{E3}) and (\ref{E4}) in Eq.\ (\ref{E2}) 
lead to the more familiar expressions
\begin{equation}
\vec{j}_{qp}(x,y)= 2\hbar m_1 \Im\left[\, \Psi^*(x,y)\, \nabla \sigma_z\, \Psi(x,y)\, \right]
+\vec{j}_{so}(x,y)\;,
\end{equation}
where
\begin{equation}
\vec{j}_{so}(x,y)=-\frac{\alpha}{\hbar}\,
\Re\left[\,\Psi^*(x,y)\,(\sigma_y \hat{x}-\sigma_x \hat{y})\tau_z\,\Psi(x,y)\,\right]\;.
\end{equation}

The charge and spin densities are obtained by adding $-e\tau_z$ and $\sigma_z$ operators, respectively, 
in Eq.\ (\ref{E2a}),
\begin{eqnarray}
\rho_{c}(x,y) &=& -e\, \Psi^*(x,y)\tau_z\Psi(x,y) \; ,\\
\rho_{s}(x,y) &=& \Psi^*(x,y)\sigma_z\Psi(x,y) \; .
\end{eqnarray}
Analogous substitutions in Eq.\ (\ref{E2}) yield the definitions of $\vec{j}_c(x,y)$
and $\vec{j}_s(x,y)$, the charge and spin currents.

The conductance of the junction is evaluated on the normal side as 
\begin{equation}
g(E)=\frac{e^2}{h}\left[\, N(E) - P_{ee}(E) + P_{eh}(E) \,\right]\; ,
\end{equation}
where 
\begin{eqnarray}
\label{eq12}
P_{ee}(E) &=&   \sum_{k,\eta_\sigma} d_{k}^{(L)}(E) \int dy \left|\phi_{k}^{(L)}(y,\eta_\sigma,\Uparrow)  \right|^2\; ,\\
\label{eq13}
P_{eh}(E) &=&  \sum_{k,\eta_\sigma} d_{k}^{(L)}(E)  \int dy  \left|  \phi_{k}^{(L)}(y,\eta_\sigma,\Downarrow)  \right|^2\; ,
\end{eqnarray}
are, respectively, the electron-electron ($ee$) and electron-hole ($eh$ or Andreev) reflection probabilities.
As well known, normal $ee$ reflection reduces the conductance while Andreev $eh$ one increases it. 
Notice also  that in the $k$-sums of Eqs.\ (\ref{eq12}) and (\ref{eq13}) only propagating output
modes have to be included. The coefficients $d_k^{(c)}$ for both evanescent and propagating modes are obtained from the numerical algorithm, with the exception of the input channels that are set to one for normalization purposes. We consider as input channels the electron propagating solutions in the normal lead with a quasi-particle flow into the junction. As a peculiarity of this problem, we found that 
for $E=0$ and $k=0$ some instabilities in the flow calculation are obtained. They are simply resolved, however, 
by using a nonzero (small)  value for $E$.  

\section{Current distributions}
\label{sec:3}

\begin{figure}
\center
\resizebox{0.75\columnwidth}{!}{%
  \includegraphics{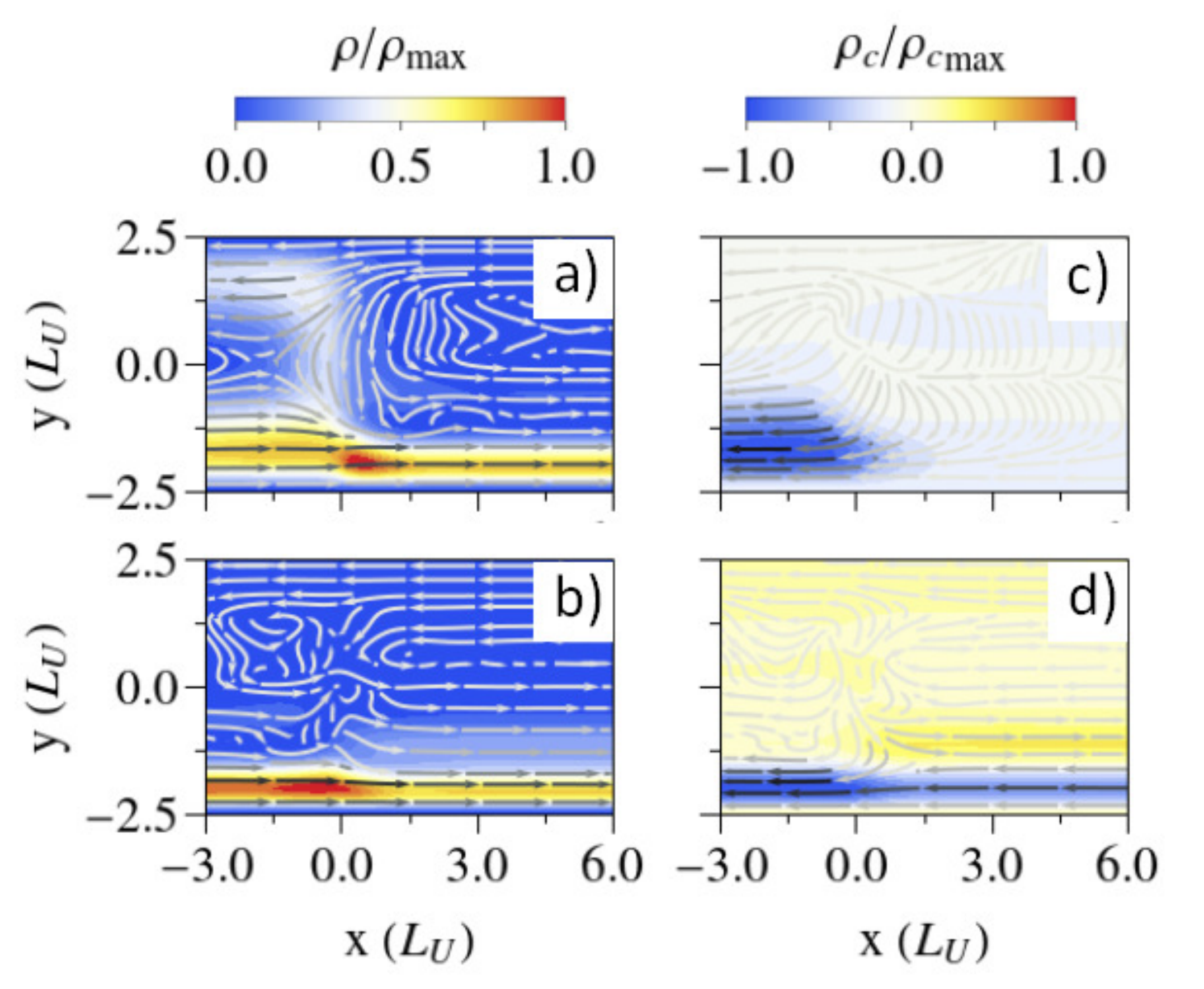}}
\caption{a) Quasi-particle current overprinted on its corresponding probability density for a NS junction with a) one chiral Majorana mode in the superconducting side of the junction (that is, ${\cal C}=1$ topological phase); b) two simultaneous Chiral Majorana modes in  the superconducting region (${\cal C}=2$ topological phase). Figures c) and d) are the charge current and densities corresponsing to the cases in a) and b), respectively. The material parameter for a) is $m_0=-1 E_U $ while for b) is $m_0=-3 E_U$. The rest of the parameters are  $\Delta=2.0 E_U$ and $E=0.1 E_U$. Note that we take $\alpha=1 E_U L_U$
and $m_1=0.5 E_U L_U^2/\hbar^2$.}
\label{F3}       
\end{figure}

In Fig.\ \ref{F3} we display the quasi-particle current distribution (arrows) overprinted on their corresponding quasiparticle densities (color or gray-shaded) for two different scenarios.
Figures \ref{F3}a and \ref{F3}c are for the case when the right side of the junction has a Chern number one, i.e.,  with a pair of topological bands crossing zero energy. Therefore, for energies below the gap  there is  a propagating Majorana mode attached to a system edge. On the other hand, Figs.\ \ref{F3}b and \ref{F3}d correspond to the case of Chern number two, with an additional pair of bands crossing zero energy. In this latter case we have simultaneously two propagating Majorana modes attached to the same edge.  The first thing we notice is that only the lower edge shows an attached Majorana channel on the right side of the junction. The reason behind this difference between upper and lower edges is that in an infinite NS junction there are no counterpropagating modes. That is, the Majorana channel in the lower border is an outgoing channel. An ingoing Majorana channel would appear on the upper edge in case we considered a second junction with a normal lead on the right of the superconductor region, with its corresponding incident modes. 

\begin{figure}
\center
\resizebox{0.5\columnwidth}{!}{%
  \includegraphics{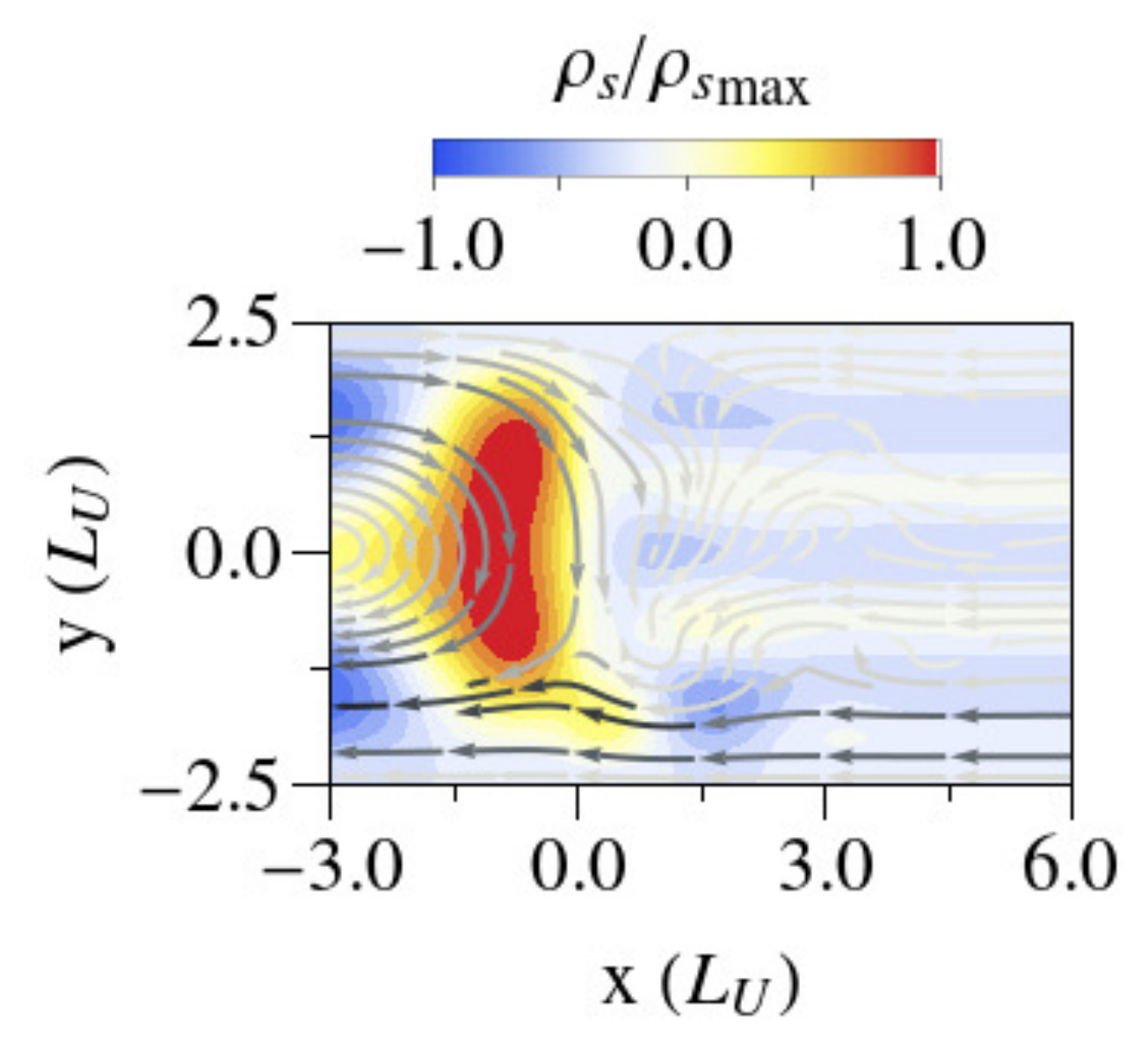}}
\caption{Spin current overprinted on the spin density for the case when the superconductor holds a single chiral Majorana mode. The Hamiltonian parameters  are $m_0=-1 E_U $, $\Delta=2 E_U$ and $E=0.1 E_U$. }
\label{F4}       
\end{figure}

As seen in Figs.\ \ref{F3}a and \ref{F3}c, with only one pair of topological bands in the superconductor region 
(${\cal C}=1$)
an incident electron channel from the normal region will be transmitted to a Majorana channel in the superconducting region. Note that the Majorana channel is associated with a zero charge density and  zero charge current. The transmission probability is $P_T=0.5$ and, nevertheless, the conductance $g(E)$ is still one quantum $g(E)=e^2/h$. The reason behind this apparent paradox is the distribution of probability between the reflected $ee$ and $eh$ channels. The electronic incident channel is partially reflected back in equal measure as an electron and as a hole through Andreev reflection, $P_{ee}=0.25$ and $P_{eh}=0.25$.
This is not in contradiction with current literature finding a conductance of $g(E)=0.5 e^2/h$ due to the Majorana mode because, as explained above, we are considering a NS junction with a single normal lead and therefore neglecting the effect in the junction from Majorana counterpropagating states with an origin in a second lead. In this sense, the reflected channels have several peculiarities. First, their charge current and densities add up to zero and the same happens with their spin current and density (see Fig.\ \ref{F4}). The incident electron channel is responsible for an ingoing spin current into the Majorana mode, signaling the topological state of the superconductor. 

On the other hand, in Figs.\ \ref{F3}b and \ref{F3}d we can see the case with two pairs of topological bands active on the right side of the junction. In this case the incident electronic channel just goes through the junction without reflection. That is not surprising because two chiral Majorana channels add up to a single electron channel. In fact, the available Majorana channels degrade with increasing energy of the incident  channel (i.e., the quality of the Majorana is worse as we deviate more and more from zero energy and approach the gap energy). In fact, we can see in Fig.\ \ref{F3}d how charge neutrality of the chiral Majoranas on the right side has been slighty lost already for $E=0.1 E_U$, probably with a certain degree of hybridization between the two Majoranas and the presence of a slight charge current in the lower superconducting border.

\begin{figure}
\center
\resizebox{0.75\columnwidth}{!}{%
  \includegraphics{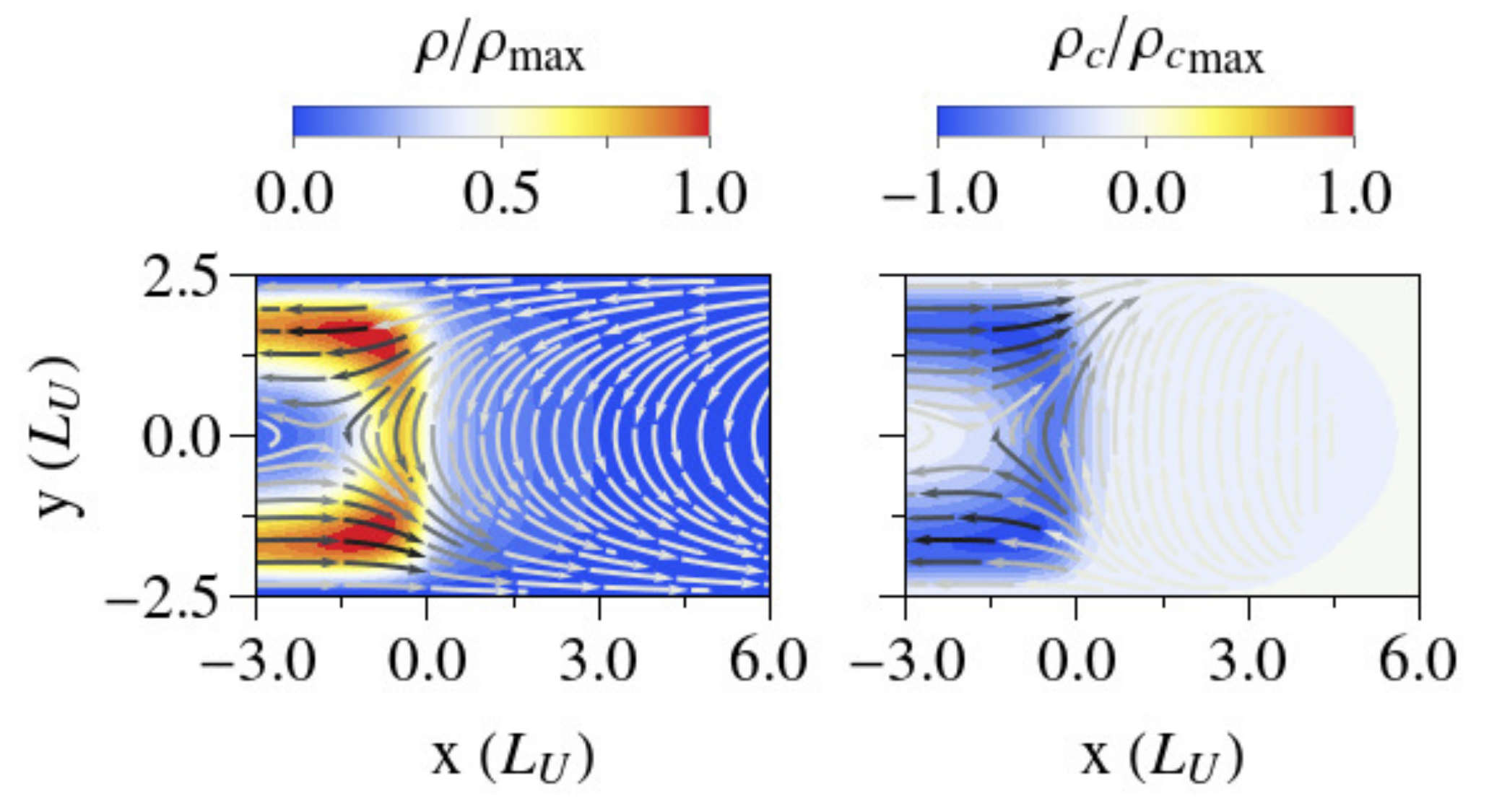}}
\caption{a) Quasi-particle current overprinted on the probability density for the case when the superconducting side of the junction is in a trivial phase. b) The same of a) for the charge current and density. In order to have open channels available to probe the superconductor, the material parameter $m_0$ takes different values on the left and right sides; it is $m_{0}=-1 E_U$  on the left and  
$m_{0}=2 E_U$ on the right. The rest of the parameters are the same of preceding figures.}
\label{F5}       
\end{figure}

In Fig.\ \ref{F5} we can see the case when the superconductor is in a trivial state. In previous figures we considered an homogeneous infinite semiconductor slab with a junction separating the proximity coupled superconducting region from the non-superconducting one. However, here for pedagogical reason we consider that the junction separates two semiconductors having different material parameter $m_0$. The reason  is that no open incident channels are available in the normal region for the range of values where the superconducting region is in a trivial phase. Therefore, we maintain the left side of the junction at a value of $m_0$ that allows for an electronic incident channel. The result is a perfect electron-electron reflection of the quasi-particle current. Therefore the overall charge and spin current in the contact remains zero.

\section{Orbital effects}
\label{sec:4}

Until now we have considered the behavior of the junction mainly regarding 
variations of the material parameter $m_0$. In the underlying physical model, this parameter relates to the magnetization of the material. In this section we want to explore how the inclusion of orbital effects due to an external magnetic field may affect the results of our model.
The strength of magnetic orbital effects is set by the magnetic length $l_z$, defined as $l_z^2=\hbar c/e B$. We consider a fully perpendicular magnetic field to the sample using a Landau gauge centered on $y=0$ through the magnetic substitution $p_x \rightarrow p_x - \hbar y/ l_z^2$. We also add the required Pauli matrix $\tau_z$ to properly consider the electron-hole symmetry of the problem \cite{Osca2015b}.
	
\begin{figure}
\center
\resizebox{1.0\columnwidth}{!}{%
  \includegraphics{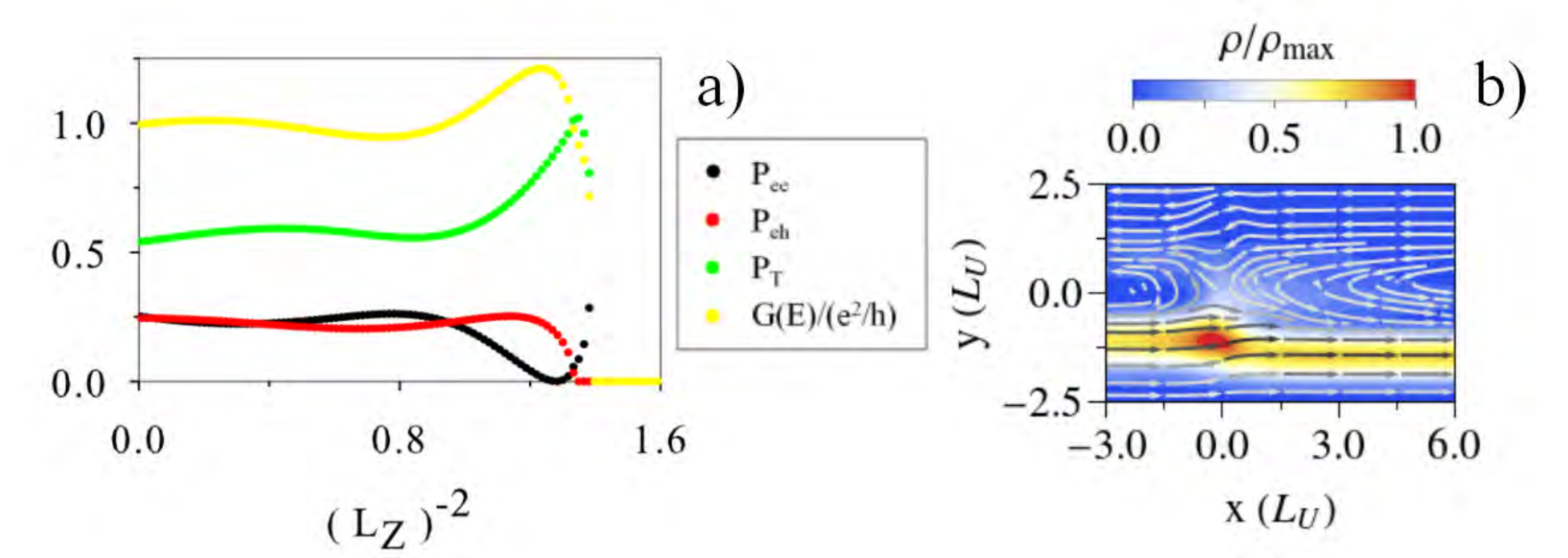}}
\caption{a) Probabilities of reflection $P_{ee}$, Andreev reflection $P_{eh}$, transmission $P_{T}$ and conductance $g(E)$ of an incident electronic channel in a QAH slab with a normal-superconductor junction. The probabilities and  conductance are shown as a function of the inverse squared magnetic length
$l_z^{-2}$ that is directly proportional to the field.
At zero field the device holds a chiral Majorana nanowire in the superconducting side of the junction. The Hamiltonian parameters  are $m_0=-1.0 E_U $, $\Delta=2.0 E_U$ and $E=0.1 E_U$. b) Quasi-particle current and probability density for $l_z^{-2}= 1.2 L_U^{-2}$. Note that, in comparison with Fig. \ref{F3}a, the position of the edge states with respect to the confinement wall has changed. There are also differences between the left and right edge states relative position in the $y$ direction.}
\label{F6}       
\end{figure}

The effects of electronic orbital motion on the QAH slab are twofold. First, if the external magnetic field is too large the edge channels disappear. This is not surprising because many chiral Majorana devices are quantum Hall devices with the addition of superconductivity. This way, different strengths of the field may enable or disable the edge propagating channels. In a certain way we are including here a competition between the QH and QAH effects. 
We can see in Fig.\ \ref{F6}a the conductance, and the different probabilities of transmission and reflection for a QAH normal-superconductor junction as a function of the magnetic length. 
At a certain value of the magnetic length ($l_z^{-2}\approx1.3 L_{U}^{-2}$) the QAH propagating channels are closed on the normal side of the junction and only evanescent modes remain. 

On the other hand, the second effect of the orbital motion is to effectively change the width of the nanowire due to magnetic confinement when $l_z<L_y$ (with $L_y$ the transverse width). This way, the distance of the QAH and chiral Majoranas with respect to the device edges increases, as can be seen comparing Fig.\ \ref{F6}b with Fig.\ \ref{F3}a. However, the most interesting feature is the separation of the propagating states from their respective edges and how this changes differently on each side of the junction for increasing external field. This affects how the electronic incident channel couples with the outgoing chiral Majorana mode on the superconductor side. Therefore, the transmission and reflection probabilities (and thus the conductance) are modified by the relative position of the channels caused by the presence of the orbital motion.	

The oscillations in reflection and transmission probabilities, and thus in conductance, are due to changes in the transversal positions of the topological states. However, these changes are abruptly hindered by the disappearance of the propagating channels in the normal lead with increasing magnetic field. In the rest of the paper we will not consider orbital effects in the normal lead of the junction, assuming that we have shielded or dampened the magnetic field in that region. This way we always have a propagating channel opened in the normal contact to probe the behavior of the chiral modes under the effects of the orbital motion.
	
\begin{figure}
\center
\resizebox{0.75\columnwidth}{!}{%
  \includegraphics{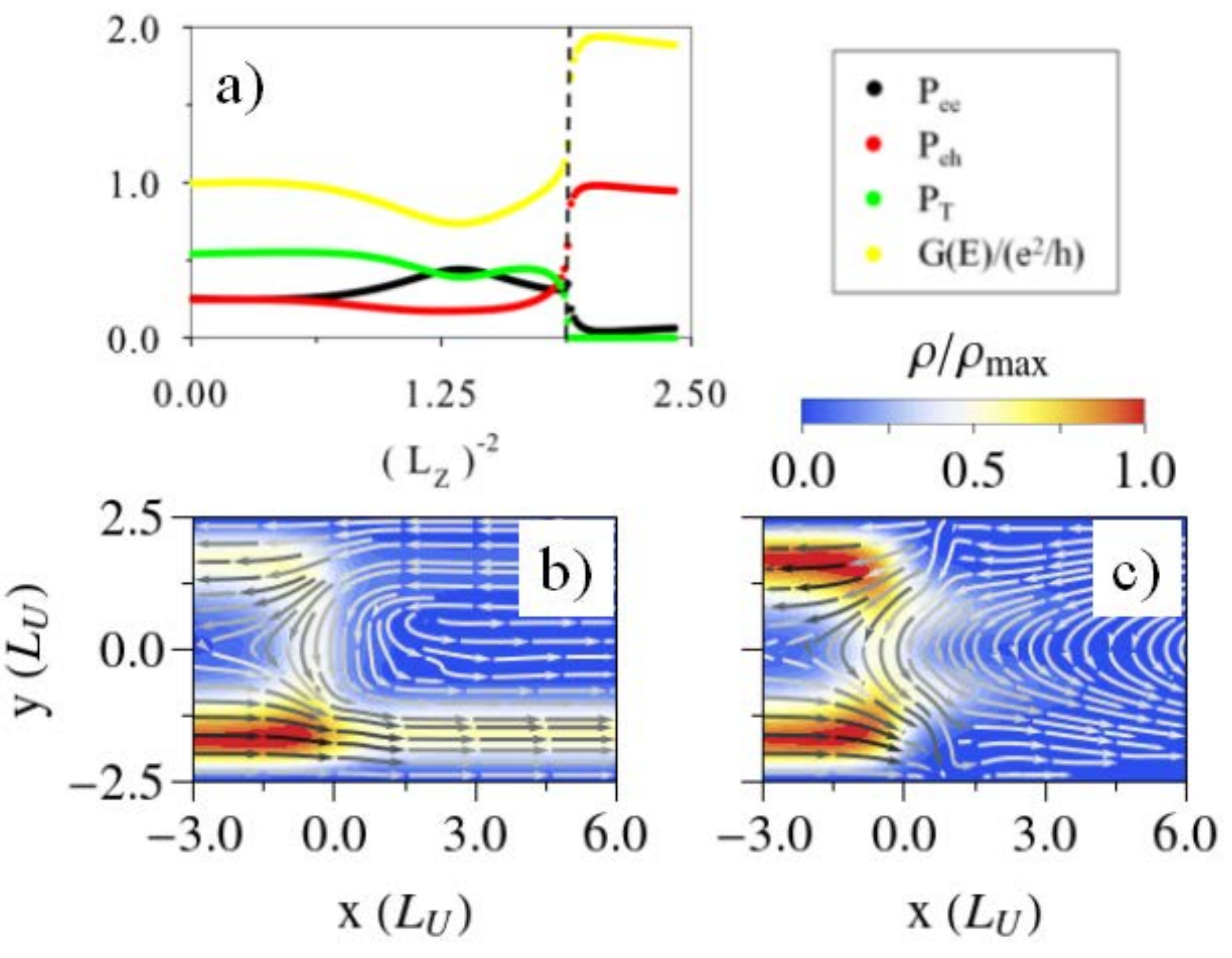}}
\caption{a) Same as in Fig.\ \ref{F6}a, but with the external magnetic field applied only to the right side of the junction. This way we avoid the channel closing on the normal side and we can probe the junction behavior for higher magnetic fields. At zero field the device holds a chiral Majorana mode in the superconducting side of the junction and the vertical dotted line signals the strength for which this Majorana mode disappears. The material parameter
$m_0=-1 E_U$ is constant all along the slab, while the rest of the Hamiltonian parameters are the same as above. b) and c) Quasi-particle current and probability density at strengths of the external field  corresponding to $l_z^{-2}= 1.2 L_U^{-2}$ and $l_z^{-2}= 2.4 L_U^{-2}$, respectively. 
Note that only evanescent modes remain on the right side in panel c). }
\label{F7}       
\end{figure}

In Fig.\ \ref{F7} we consider a QAH slab with orbital effects active only on the superconducting side. The superconducting region is tuned to hold a single Majorana channel at zero external field. We can see in Fig.\ \ref{F7}a (at the left of the vertical dashed line) how the transmission probability slightly decreases while the normal reflection increases with increasing magnetic strength. The reason is the change in spatial alignment between the incident and the Majorana channels, as shown in Fig.\ \ref{F7}b. This behavior 
persists up to the strength value marked as a black vertical dashed line. From that point onwards the magnetic effective confinement is too narrow to allow the nanowire to hold the transversal length of the Majorana. Therefore the propagating chiral Majorana mode disappears and only evanescent modes remain in the superconducting region. This is signaled by a zero transmission probability and the dominance of the Andreev effect as the main reflection mechanism. Electron-hole reflection probability rises to one and  the conductance achieves its maximum value of two.  

\begin{figure}
\center
\resizebox{0.75\columnwidth}{!}{%
  \includegraphics{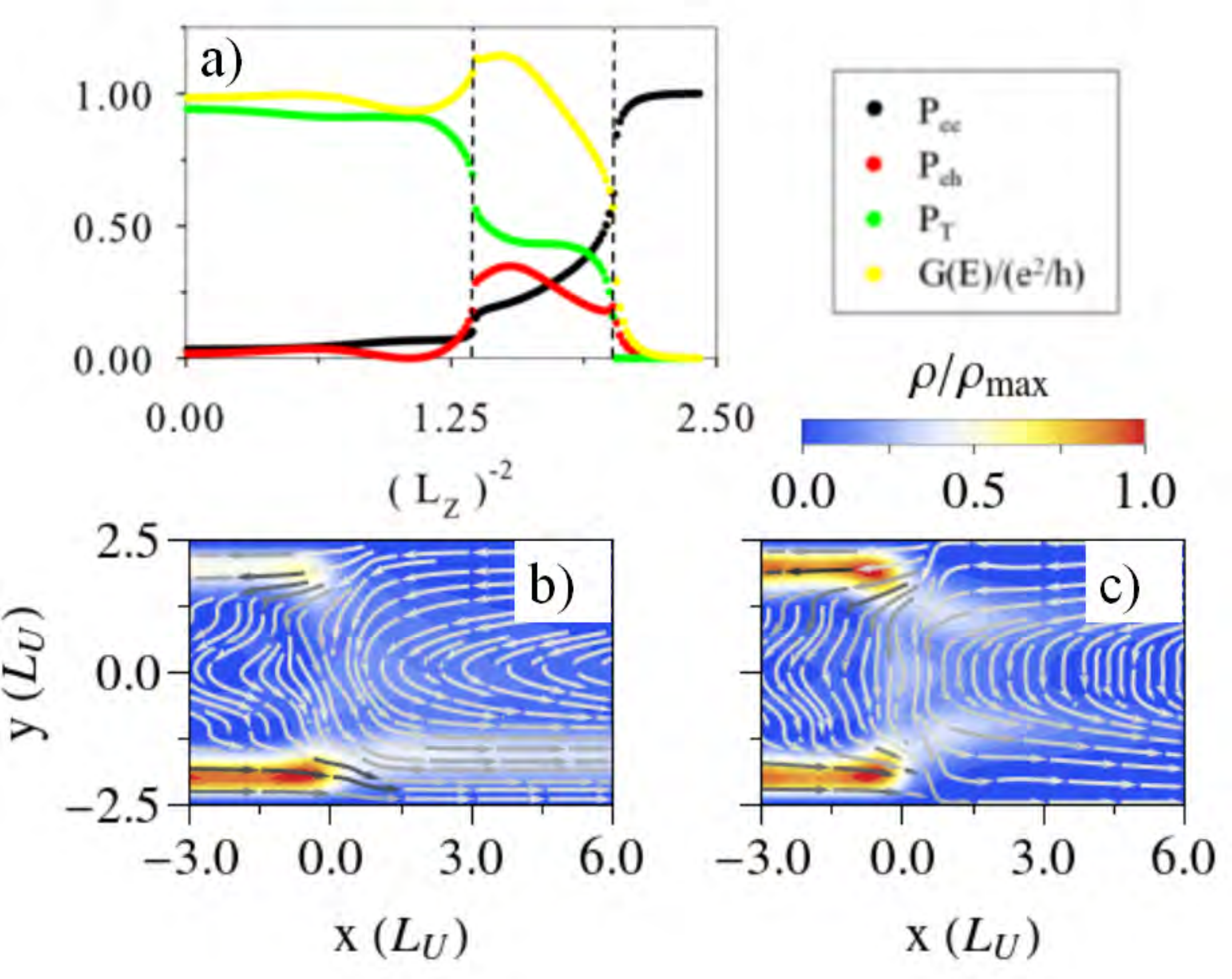}}
\caption{a) Same as in Fig.\ \ref{F7}a but with a material parameter $m_0=-3 E_U$. The rest of the Hamiltonian parameters are the same as above. This way, at zero field the device holds two chiral Majorana modes in the superconducting side of the junction. Each vertical dotted line in a) signals the strength for which one Majorana mode is lost.  b) and c) Quasi-particle current and probability density for strengths of the external field at the right side of the junction 
corresponding to $l_z^{-2}=1.6 L_U^{-2}$ and $l_z^{-2}=2.4 L_U^{-2}$, respectively.}
\label{F8}       
\end{figure}

Finally, in Fig.\ \ref{F8} we consider the same slab but with the superconducting region tuned to hold two Majorana channels at  zero external field. In Fig.\ \ref{F8}a the first vertical dashed line signals the transition from a state with two Majorana edge states to a single Majorana state, while the second one signals the loss of both Majorana channels. The first transition is followed by a change in the transmission probability $P_T\approx 1$ to $P_T\approx 0.5$ as we expect from the loss of one of the two Majorana channels.  Accordingly, the electron and hole reflection probabilities rise from zero to $P_{ee}\approx P_{eh}\approx0.25$. Note, however, that here the change of the probabilities with the magnetic strength is not abrupt (probably because of large transverse finite size effects). The change is also smooth at the transition from one to zero active Majorana channels. This causes the conductance to oscillate while the system evolves between different conductance plateaus with smooth oscillations. 
	
\section{Conclusion}
\label{sec:5}
We have studied how the conductance in an normal-superconductor junction with chiral Majorana modes is related to the spatial distribution of currents using a simplified model. In particular, we have shown how the spatial coupling of the propagating modes on the different sides of the junction is relevant to explain the observed results. Furthermore, we have introduced the effect of the orbital motion in the model to investigate how this coupling is affected by a magnetic field. It is the objective of future work to apply this type of analysis to a more realistic physical model, like that of Ref.\ \cite{He294}, where we expect to observe similar behaviors plus some additional ones. The reason is that many of these models may be rewritten in terms of one or several coupled
copies of the present one. 

\section*{Acknowledgments}
This work was funded by MINEICO-Spain, grant MAT2017-82639.

\bibliographystyle{epjc}
\bibliography{circdictt_noURL}

\end{document}